\documentclass[a4paper,11pt]{article}
\usepackage{jinstpub} 
\usepackage{orcidlink}

\usepackage{lineno}
\usepackage{threeparttable}
\usepackage{booktabs}
\usepackage{multirow}
\usepackage{dcolumn}
\usepackage{arydshln}
\usepackage{upgreek}
\usepackage{extarrows}
\usepackage{amsfonts}
\usepackage{wasysym}
\usepackage{csquotes}
\usepackage{booktabs}
\usepackage{multirow}
\usepackage{threeparttable}
\usepackage{dcolumn}
\usepackage{colortbl}
\usepackage[nameinlink,capitalize]{cleveref}

\newcommand{\isotope}[2]{\ensuremath{{}^{\mathrm{#1}}\mathrm{#2}}~}
\newcommand{\decay}[2]{\xLongrightarrow[\mathrm{{#2}}]{\mathrm{{#1}\,MeV}}}
\newcommand{\decayb}[2]{\xlongrightarrow[\mathrm{{#2}}]{\mathrm{Q_\upbeta = #1\,MeV}}}

\title{Characterization of a $^\mathbf{220}$Rn source for low-energy electronic recoil calibration of the XENONnT detector}

\author[a,1]{Florian J\"org\orcidlink{0000-0003-1719-3294},\note{Corresponding author.}}
\author[b,1,2]{Shengchao Li\orcidlink{0000-0003-0379-1111},\note{Now at Westlake University}}
\author[a]{Jochen Schreiner,}
\author[a]{Hardy Simgen\orcidlink{0000-0003-3074-0395},}
\author[b]{Rafael F. Lang\orcidlink{0000-0001-7594-2746}}

\affiliation[a]{Max-Planck-Institut f\"ur Kernphysik\\ Saupfercheckweg 1, 69117 Heidelberg, Germany}
\affiliation[b]{Department of Physics and Astronomy, Purdue University\\ 525 Northwestern Ave, West Lafayette, IN 47907, USA}

\emailAdd{fjoerg@mpi-hd.mpg.de}
\emailAdd{li4006@purdue.edu}

\abstract{
Low-background liquid xenon detectors are utilized in the investigation of rare events, including dark matter and neutrinoless double beta decay. For their calibration, gaseous $^{220}$Rn can be used. After being introduced into the xenon, its progeny isotope $^{212}$Pb induces homogeneously distributed, low-energy ($<30$\,keV) electronic recoil interactions. We report on the characterization of such a source for use in the XENONnT experiment. It consists of four commercially available $^{228}$Th sources with an activity of 55\,kBq. These sources provide a high $^{220}$Rn emanation rate of about 8\,kBq. We find no indication for the release of the long-lived $^{228}$Th above 1.7\,mBq. Though an unexpected $^{222}$Rn emanation rate of about 3.6\,mBq is observed, this source is still in line with the requirements for the XENONnT experiment.
}


\begin{document}
\maketitle
\flushbottom
\section{Introduction}\label{sec:intro}

Detectors utilizing liquid xenon to search for neutrinos and rare events have undergone rapid growth in both mass and radiopurity during the past decade~\cite{XENONCollaboration:2022kmb,LZ:2022ufs,PandaX-4T:2021bab,EXO-200:2019rkq}. The time projection chambers (TPCs) used in these experiments utilized the high-Z property of the xenon target to shield the fiducial volume from surrounding radioactivity. However, this means any external gamma calibration source will not reach the center of the TPC, as the mean free path for example of MeV gammas is of $\mathcal{O}$(cm) and thus much smaller than the diameter of the meter-scale TPCs.

In this paper, we study an internal source to calibrate the low-energy (keV-scale) electronic recoil response of the XENONnT TPC~\cite{XENON:2020kmp, XENON:2023}. $^{220}$Rn emanated from $^{228}$Th is carried by the xenon gas flow and mixed with the liquid target. The $^{220}$Rn decay chain produces a variety of radiation:
\begin{align}
&\isotope{228}{Th} \decay{5.5}{1.9\,y} \isotope{224}{Ra} \decay{5.8}{3.6\,d} \isotope{220}{Rn} \decay{6.4}{56\,s} \isotope{216}{Po} \decay{6.9}{145\,ms} \isotope{212}{Pb}\nonumber\\
&\decayb{0.6}{11\,hrs} \isotope{212}{Bi}
\begin{array}{l}
{}^{36\%}\nearrow~\decay{6.2}{61\,min} \isotope{208}{Tl} \decayb{5.0}{3.1\,min}~\searrow\\
\\
{}_{64\%}\searrow~\decayb{2.3}{61\,min} \isotope{212}{Po} \decay{9.0}{294\,ns}~\nearrow
\end{array}
\isotope{208}{Pb} 
\qquad
\boxed{
\begin{array}{l}
\xLongrightarrow{\mathrm{\upalpha-decay}}\\
\xlongrightarrow{\mathrm{\upbeta-decay}}
\end{array}
}
\label{eq:228_th_decay_chain}
\end{align}
In the $^{228}$Th decay chain, $^{212}$Pb is relatively long-lived with a half-life of 10.6~h, providing a beta-spectrum with a uniform component below 200~keV electronic recoil energy, and a high event rate above this energy. This calibrates the major background from the beta emitter $^{214}$Pb originating from $^{222}$Rn, which constitutes about 50\% of the electronic recoil~(ER) background below 10\,keV in the first science data set of the XENONnT~\cite{XENONCollaboration:2022kmb} experiment. Such a low-energy calibration source is essential to estimate the electronic recoil background contributing to the region of interest to the dark matter search~\cite{XENON:2018voc}. Following tests in a small detector~\cite{Lang:2016zde}, the first calibration using a $^{220}$Rn source was implemented in XENON100~\cite{XENON:2016rze}. As the active mass of dark matter detectors has since increased by 2~orders of magnitude, a higher radon emanation rate is desired to produce the required statistics for calibration; meanwhile, the average ER background rate (events per keV$\times$tonne$\times$year) has decreased by 2~orders of magnitude, which places a stringent limit on the radiopurity of the calibration source with a more precise radioactive background estimation. These requirements for the source are crucial for applications for low-background measurements.

The design of the $^{228}$Th source is described in \autoref{sec:source}, while measurements of its $^{220}$Rn and $^{222}$Rn emanation rate are reported on in \autoref{sec:rn220_emanation} and \autoref{sec:rn222_emanation}.
To prevent contamination of the experiment with the long-lived $^{228}$Th, its release has been excluded by a measurement which is presented in \autoref{sec:thoron_residue}.
Taken together, these measurements confirm that this source meets or exceeds the requirements of a low-background experiment such as XENONnT.

\section{Source Preparation}\label{sec:preparation} \label{sec:source}
\begin{figure}[h]
    \centering
    \includegraphics[width=0.9\textwidth]{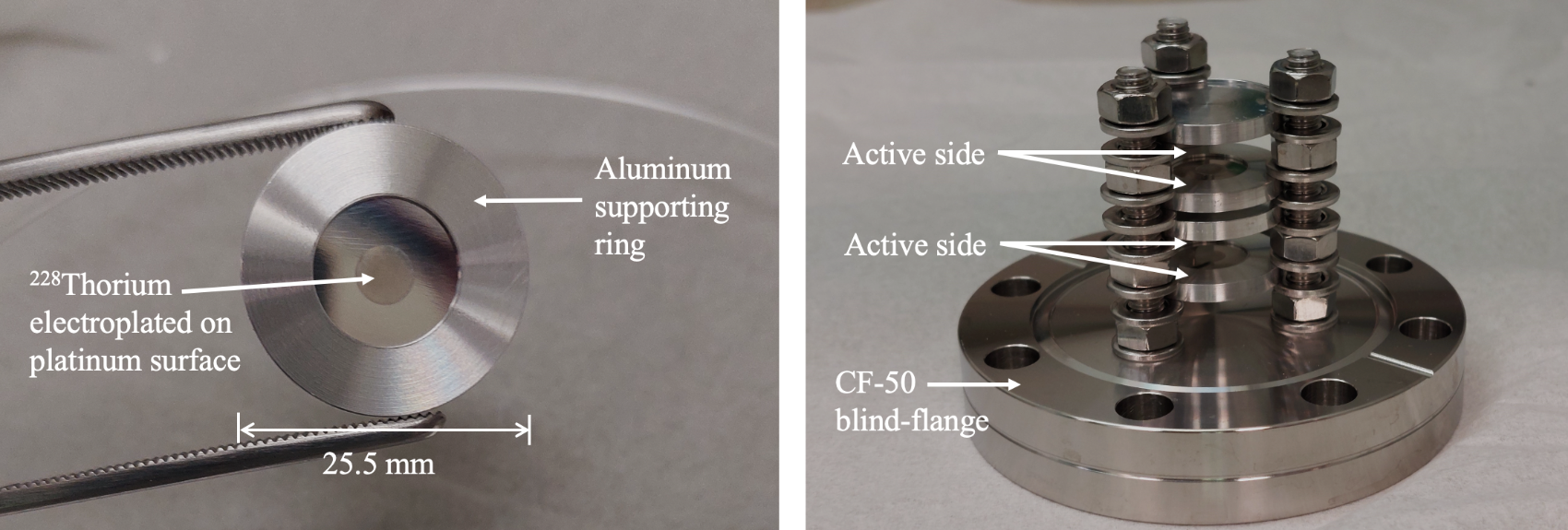}
    \caption{\textbf{Left:} Individual 25.5\,mm-diameter platinum disc with $^{228}$Th oxide deposition visible in the center. \textbf{Right:} Final assembly of the four source discs on a CF-50 blind flange using M6 threaded rods and nuts. The active sides of the sources are facing each other with spaces to maximize the radon emanation.}
    \label{fig:th228_source}
\end{figure}

Four $^{228}$Th source discs were procured from Eckert \& Ziegler~\cite{EZAG}, with a nominal activity of $\sim$13.8\,kBq each, as of April 2022. The $^{228}$Th oxide was electroplated to a 5\,mm-diameter active area at the center of each 25.5\,mm-diameter platinum disc. The usual gold covering on the active area was not added to enhance radon emanation. The four discs were placed in an emanation vessel with Conflat\textsuperscript{\textregistered} CF-50 stainless steel (SS) flanges. The source was held by oversized washers and nuts that were mounted on three M6 SS threaded rods that were attached to a CF-50 blind flange. 
The discs were mounted with the active sides facing each other, as shown in \autoref{fig:th228_source}. To minimize contamination, all parts of the support structure were thoroughly cleaned prior to assembly according to the procedure outlined in\,\cite{XENON:2021mrg}. 

\section{Measurement of the $^{220}$Rn emanation rate}\label{sec:rn220_emanation}

\begin{figure}[h]
    \centering
    \includegraphics[width=0.5\textwidth]{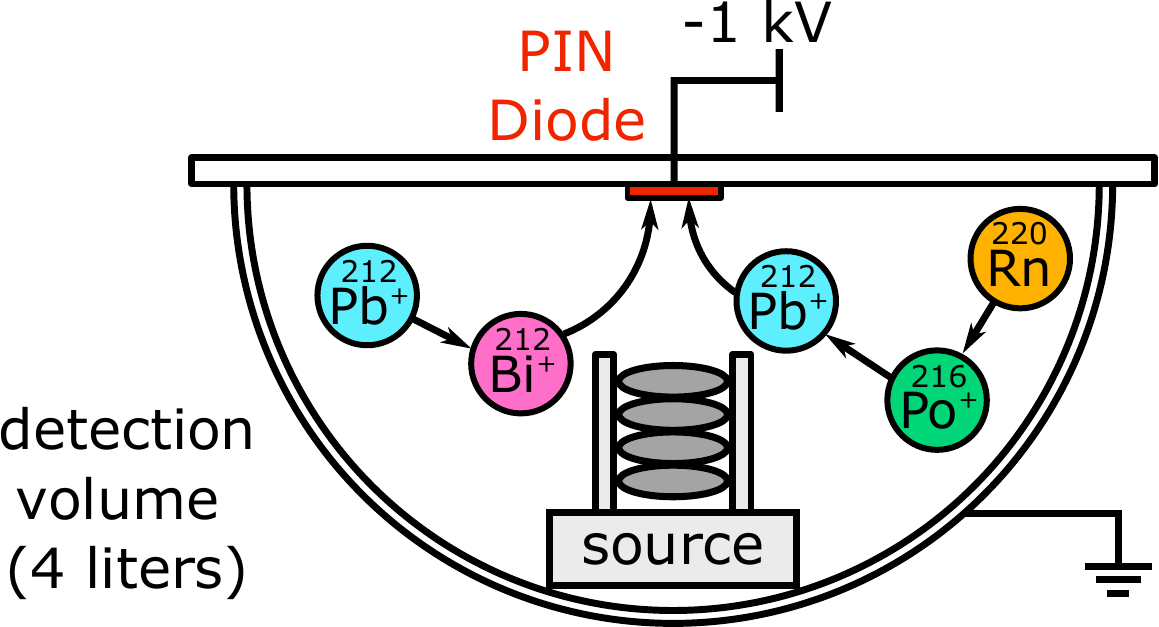} 
    \caption{A schematic drawing of the electrostatic radon monitor. Positively charged radon progeny particles are attracted by an electric field toward the Si PIN diode, which acts as an alpha-detector. Figure adapted from~\cite{Joerg2017}.
}
    \label{fig:rn_monitor}
\end{figure}

\begin{figure}[h]
    \centering
    \includegraphics[width=0.8\textwidth]{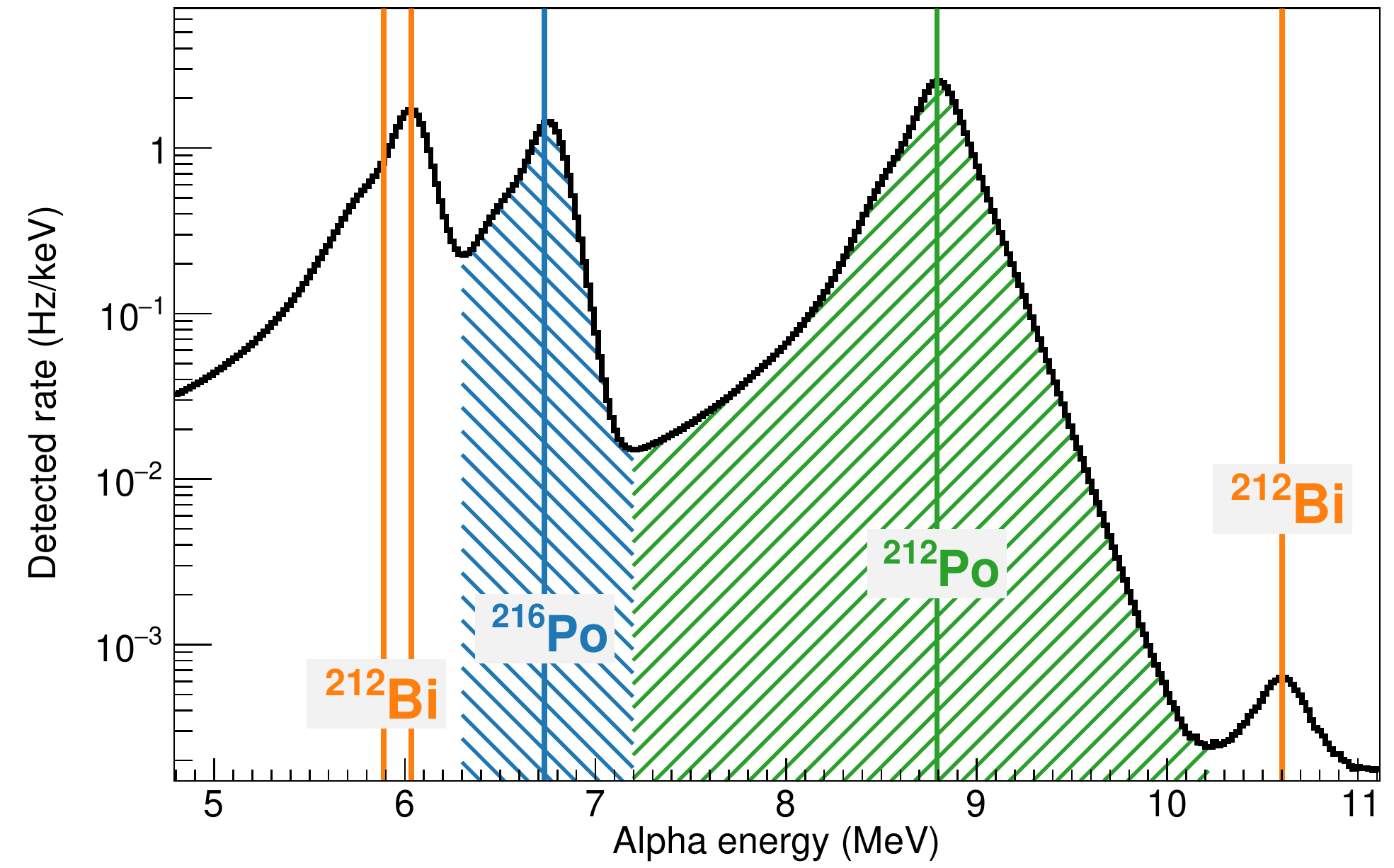}%
    \caption{Energy spectrum of electrostatically collected $^{220}$Rn progeny from the source, measured in the electrostatic radon monitor (black) over the four-day-long measurement period. Note that the emission line of $^{220}$Rn itself is not visible, as the detector relies on the collection of charged radon progeny.}
    \label{fig:alpha_spectrum}
\end{figure}

During the measurement, the assembly shown in the right of \autoref{fig:th228_source} was placed directly inside the detection volume of the electrostatic radon monitor~\cite{Brunner:2017xsu}, whose working principle is shown in \autoref{fig:rn_monitor}.
The four-liter hemispheric detector volume is filled with nitrogen at a slight over-pressure of 1050\,mbar and is instrumented with a Si-PIN diode mounted to the top flange.
By applying a -1\,kV bias voltage, the positively charged radon progeny ions are collected at the surface of the diode.
Alpha particles emitted during their subsequent decays can then be detected, in case they are directed towards the diode.
\autoref{fig:alpha_spectrum} shows the alpha-particle energy spectrum of the collected radon daughters, with clear emission lines from $^{216}$Po, $^{212}$Bi and $^{212}$Po, as expected from the $^{220}$Rn chain (\autoref{eq:228_th_decay_chain}).
While the observed alpha rates match the expectation from the known $^{212}$Bi branching ratio, the rate of $^{216}$Po (blue hatched region) is found to be suppressed.
This is a consequence of the isotope's short half-life (145\,ms), which is of the same order of magnitude as the median duration of the collection process.
Therefore, a certain fraction of the $^{216}$Po ions disintegrates along their way, before reaching the Si-PIN diode.
Note that this does not affect the even shorter lived $^{212}$Po, as its progenitor isotopes are already collected on the diode.
The subdominant peak at 10.8\,MeV is compatible with the rare \enquote{long-range alpha} transitions between $^{212}$Bi and the ground state of $^{208}$Pb that occurs in 0.014\% of the $^{212}$Bi alpha decays\,\cite{Rytz:1951,Bertolini:1962,leang1965}.

To prevent the uncertainty of the $^{216}$Po rate due to its in-drift-decay, the $^{220}$Rn emanation rate of the source is determined solely from the equilibrium rate of $^{212}$Po decays.
They are selected within the green hatched area shown in \autoref{fig:alpha_spectrum}.
The fraction of events falling outside this selection is estimated from a fit to spectrum and amounts to 2\%.
Due to the 11 hour-long half-life of the preceding $^{212}$Pb isotope, the activity evolution of $^{212}$Po features a time-delayed increase towards its equilibrium value.
Therefore, its initial activity $A_\text{init}\big({}^{212}\text{Po}\big)$ that is determined from a time interval $[t_1, t_2]$ at the beginning of the measurement, needs to be up-scaled by the fraction $f_\text{eq}\big({}^{212}\text{Po}\big)$ to which its equilibrium activity has been reached (see \autoref{eq:rn220_conversion}).
We determine this fraction from an analytical calculation\,\cite{moral2003} taking into account the complete decay chain dynamics.
To reduce the impact of local charge accumulation caused by the high activity of the source, a time interval of approximately 6 hours ($t_1=20\,\text{min}, t_2=240\,\text{min}$) was chosen from the four-day-long measurement.
Due to the steep increase of the $^{212}$Po rate at the beginning of the measurement, a delay between the start of the radon progeny collection and the beginning of data acquisition would lead to a variation of up to 8\% of the final result.
This was estimated by shifting the selected time-window $[t_1, t_2]$ by $\pm$5 minutes in either direction.

Finally, the total detection efficiency for $^{220}$Rn has to be taken into account.
Since a $^{220}$Rn reference source was unavailable, it could not be measured directly.
Therefore, the approach that has been applied in\,\cite{Lang:2016zde} was followed, and the efficiency was estimated using the one for the radon isotope $^{222}$Rn.
The latter has been measured for our detector under similar conditions to be $\epsilon\big({}^{222}\mathrm{Rn}\,\big|\,{}^{214}\mathrm{Po}\big) = (35\pm2)\%$\,\cite{Jorg:2022tli,Jorg:2022spz}, when using the detected rate of the $^{222}$Rn progenitor isotope $^{214}$Po exclusively.
Since $^{212}$Po and $^{214}$Po appear on the same location along their respective decay chain, their collection probability should be very similar.
After accounting for the 64\% branching fraction of $^{212}$Bi (see decay chain in \autoref{eq:228_th_decay_chain}), a detection efficiency of $\epsilon\big({}^{220}\mathrm{Rn}\,\big|\,{}^{212}\mathrm{Po}\big) = (22.4 \pm 1.3)\%$ for $^{220}$Rn can be assumed.
We attribute an additional uncertainty of 1.4\% accounting for field-free regions generated by the presence of the $^{220}$Rn source (e.g. below the source assembly), from which no ions can be collected.

The $^{220}$Rn emanation rate $R\big(\mathrm{^{220}Rn}\big)$ of the source is then given by
\begin{align}
    R\big(\mathrm{^{220}Rn}\big) = \frac{1}{\epsilon\big({}^{220}\mathrm{Rn}\,\big|\,{}^{212}\mathrm{Po}\big)}\cdot
    \frac{A_\text{init}\big({}^{212}\text{Po}\big)}{f_\text{eq}\big({}^{212}\text{Po}\big)}
    \,,\label{eq:rn220_conversion}
\end{align}
and a value of $R\big(\mathrm{^{220}Rn}\big) = (8.2\pm0.8)\,\mathrm{kBq}$ is obtained.
The uncertainty corresponds to the squared sum of the individual contributions discussed above, with the statistical uncertainty being found to be negligible.
The measured radon emanation efficiency is 15\%, which denotes the fraction of the $^{220}$Rn emanated from the source over the total amount produced by the decay of $^{228}$Th. 
This result is compatible to the one of the radon source in Ref.~\cite{Chott:2022lnc}, 3 times more than the source produced by PTB (Germany) for XENON1T~\cite{Lang:2016zde}, and 3 orders of magnitude higher than natural thorium compound sources~\cite{Ma:2020kll}. 
This improvement is attributed to the increased surface area, achieved by distributing the total activity among four individual discs, by which a more efficient radon release is obtained.

\section{$^{222}$Rn emanation of the source}\label{sec:rn222_emanation}

Due to the long half-life of $^{222}$Rn ($\mathrm{T_{1/2}}=3.8$\,days), an excessive emanation of this isotope to the XENONnT detector must be prevented.
To measure the $^{222}$Rn emanation rate, the source vessel has been filled with helium and was left for accumulation of the radon for several days. Then the gas was transferred into an evacuated 20\,liter expansion vessel, in which the short-lived $^{220}$Rn was left to decay to a negligible level over a time of 4.5\,hours.
The remaining $^{222}$Rn was then extracted, purified and collected by an activated carbon trap that is cooled to liquid nitrogen temperature\,\cite{Aprile:2020rn}.

This radon is subsequently transferred along with a counting gas mixture consisting of 90\% argon and 10\% methane (P10) into a miniaturized proportional counter\,\cite{Aprile:2020rn, Zuzel:2009}.
Although these highly radio-pure counters do not allow for a full spectroscopic reconstruction of the radon alpha spectrum, alpha decays can be clearly separated from background events induced by muons, beta- or gamma-rays. 
Thus, we use these detectors as simple counters with a fixed energy threshold of 50\,keV. 
We calibrate the total $^{222}$Rn transfer and counting efficiency with an acidic $^{226}$Ra standard solution producing a well known amount of $^{222}$Rn. 
From all alpha decays of the $^{222}$Rn decay chain, ($1.48\pm0.06$) alpha events are recorded for each radon atom, on average\,\cite{Aprile:2020rn}.

Two such measurements were performed, consistently revealing a $^{222}$Rn emanation rate of $(3.62 \pm 0.14)$\,mBq.
Given that the previous $^{228}$Th source produced by the PTB (Germany) did not show any significant $^{222}$Rn emanation above $<50\,\upmu$Bq\,\cite{Lang:2016zde}, this finding was unexpected. 
Given that Ref.~\cite{Chott:2022lnc} reported a comparable $^{222}$Rn emanation from a similar source manufactured by Eckert \& Ziegler, it is likely that the $^{222}$Rn is released by trace impurities introduced during the production process.
Under the assumption that both radon isotopes have the same emanation fraction, the level of $^{226}$Ra impurity in the $^{228}$Th deposit can be estimated to be 350\,ppm.

While the $^{222}$Rn emanation of this source corresponds to about 10\% of the total $^{222}$Rn emanation of the XENONnT experiment\,\cite{XENON:2021mrg}, its effect is expected to become subdominant approximately one week after the end of the $^{220}$Rn calibration.
This waiting time is further reduced by the radon removal system of the XENONnT experiment\,\cite{Murra:2022mlr} and will thus not appreciably impact the data-taking efficiency.
For future large-scale experiments, such as DARWIN/XLZD\,\cite{DARWIN:2016hyl, Aalbers:2022dzr} and nEXO\,\cite{nEXO:2021ujk}, however, this might not necessarily be true due to their more demanding background requirements.
Therefore, it will be worth investigating possibilities to identify the origin of this impurity in order to mitigate it in future source productions.

\section{Exclusion of $^{228}$Th release}\label{sec:thoron_residue}

\begin{figure}[h]
    \centering
    \includegraphics[width=0.9\textwidth]{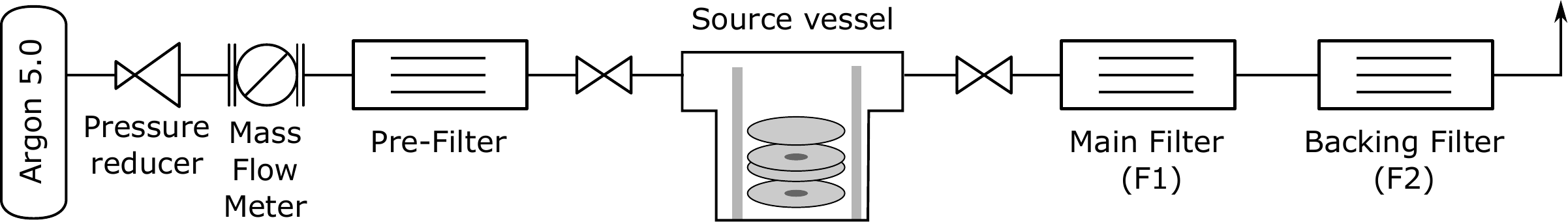}%
    \caption{Sketch of the setup used to certify that no $^{228}$Th is removed from the source. Any potential $^{228}$Th removed from the source is carried with the argon purge and collected on either of the two downstream filters (F1\&F2). After 9\,days of continuous flushing, the activity on both filters is measured using high-purity germanium (HPGe) spectrometers.}
    \label{fig:flush_through_test}
\end{figure}

Contamination due to $^{228}$Th and its decay products must be avoided in all components in the XENONnT experiment. Therefore, it must be guaranteed that no thorium is released from the source.
To test this, the source was flushed for 9\,days with argon at a flow rate of 700\,SCCM using the setup sketched in \autoref{fig:flush_through_test}.
Non-gaseous components in the argon stream are collected by two 0.2\,$\upmu$m PTFE membrane filters.
To check for potential contamination with $^{228}$Th, both downstream filters (F1 \& F2) were then measured using high purity germanium (HPGe) spectrometers located underground at a shallow depth of 15 meters of water equivalent\,\cite{Laubenstein:2004}.
The activity on both filters was evaluated via the $^{228}$Th progeny $^{212}$Pb. 
For this a weighted mean of its own gamma emission line at 238.6\,keV as well as two lines of its progeny isotope $^{208}$Tl (583.2\,keV \& 2614.5\,keV) was used, which is referred to as the $^{212}$Pb activity in the following.

\begin{figure}[h]
    \centering
    \includegraphics[width=0.85\textwidth]{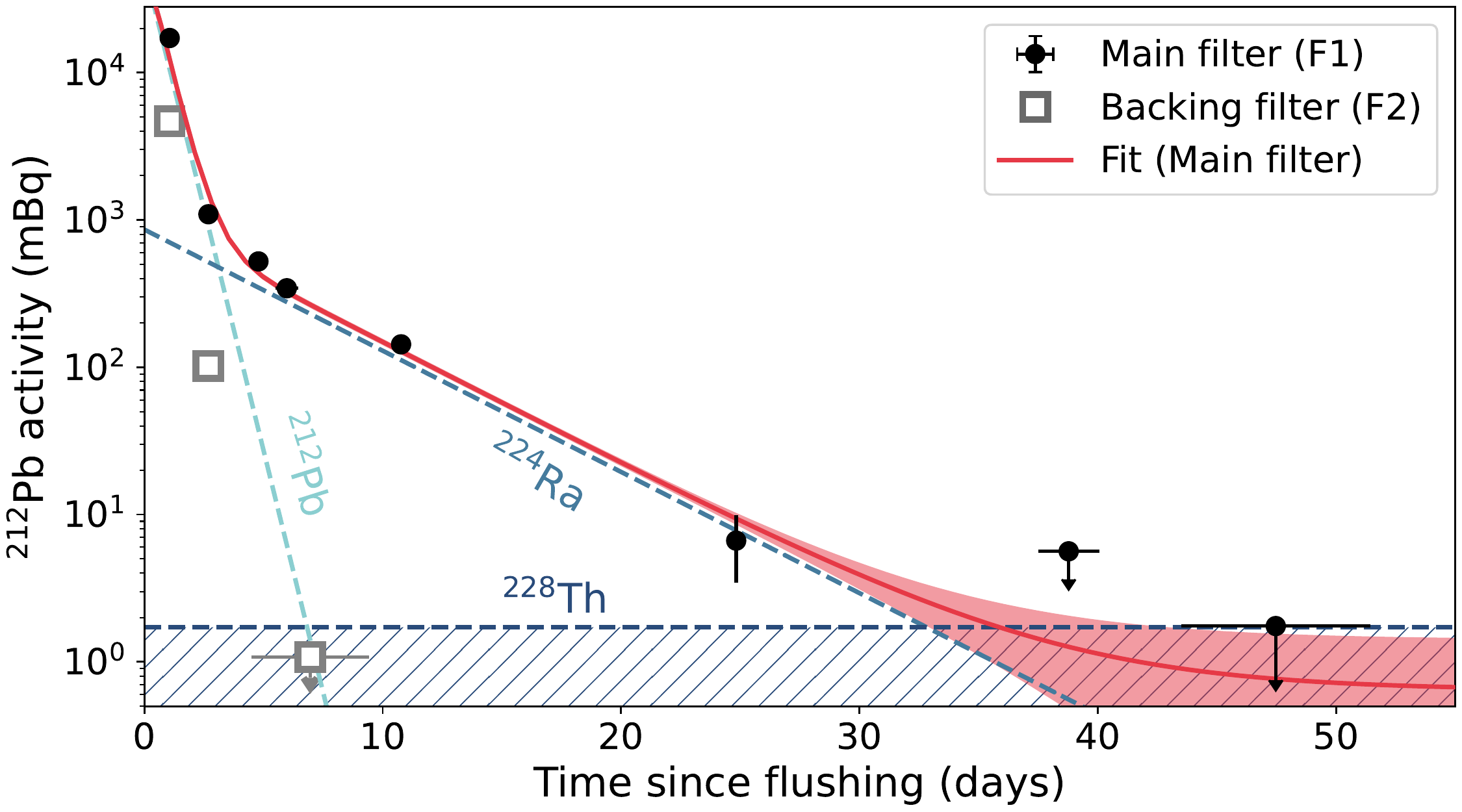}%
    \caption{HPGe measurement of the $^{212}$Pb on the main filter (F1, black dots) and backing filter (F2, gray squares). The activity evolution on F1 is well described by the radioactive decay chain (\autoref{eq:228_th_decay_chain}) with different initial $^{212}$Pb and $^{224}$Ra activities. The red line shows a fit of the activity  on F1 together with its 68\% credible region (red band). The contributions of $^{212}$Pb and $^{224}$Ra to the time evolution of the activity are indicated by the dashed blue lines.
    No contribution from the long-lived $^{228}$Th isotope is found above the upper limit indicated by the dark-blue hatched area.
    }
    \label{fig:result_flush_through_test}
\end{figure}

The time evolution of their activity is shown in \autoref{fig:result_flush_through_test}.
The activity on the main filter (F1) is initially dominated by $^{212}$Pb accumulated in the plate-out of $^{220}$Rn progeny.
This contribution quickly decays with a half-life of about 11\,hours (dashed light blue line in \autoref{fig:result_flush_through_test}).
Afterward, the evolution follows the decay of $^{224}$Ra, which is released via recoil from the source and has a half-life of 3.6\,days (dashed medium blue line).
Finally, a contribution from $^{228}$Th would cause a persisting, constant activity.
As there is no detectable activity after 42\,days, an upper limit of $\leq 1.7$\,mBq at 90\% C.L. can be placed as indicated by the hatched dark blue area.
This is sufficient given the requirements of the XENONnT experiment.
The activity evolution of the backing filter (F2) is found to follow the decay of $^{212}$Pb without any indication for $^{224}$Ra.
This indicates that all non-noble gas components have been collected already by the main filter (F1).
Only the $^{220}$Rn progeny produced in between F1 and F2 are then collected on the backing filter. 

\section{Summary}\label{sec:conclusion}

Internal calibration sources are an essential tool to calibrate large-scale liquid xenon detectors used for rare event searches.
In this work, we have characterized a new $^{228}$Th source, which has been thereafter installed and successfully applied in the XENONnT experiment.
This source produces gaseous $^{220}$Rn, that can be introduced and mixed into the liquid xenon and allows for a homogeneous calibration via the low energy electronic recoil signals from the decay of $^{212}$Pb.

\begin{table}[h]
    \centering
    \begin{tabular}{lc}
    \toprule
         Measurement          &    Result  \\
         \midrule
         $^{228}$Th activity  &  4 $\times$ 13.8\,kBq\\  
         $^{220}$Rn emanation & $(8.2\pm0.8)$\,kBq \\
         $^{222}$Rn emanation & $(3.62 \pm 0.14)$\,mBq\\
         $^{228}$Th release   & $\leq$ 1.7 mBq (90\% C.L.)\\
         \bottomrule
    \end{tabular}
    \caption{Total $^{228}$Th activity of the new XENONnT source as provided by the supplier with results from the characterization measurements carried out in this work. The elevated $^{222}$Rn emanation is likely caused by trace impurities introduced during source manufacturing.}
    \label{tab:summary}
\end{table}

The new source consists of four commercially available $^{228}$Th-plated discs (see \autoref{fig:th228_source}).
This design showed an enhancement of the $^{220}$Rn emanation rate as compared to similar sources applied in other experiments\,\cite{Lang:2016zde, Ma:2020kll}.
Though an unexpectedly large $^{222}$Rn emanation rate has been found, the source is still compatible with an application in the XENONnT experiment.
Furthermore, the release of the long-lived $^{228}$Th from the source was excluded experimentally.
We summarize the results of our measurements in \autoref{tab:summary}. 
Gaseous sources providing $^{220}$Rn emanation will be an important tool also for the next generation of liquid xenon detectors such as nEXO\,\cite{nEXO:2021ujk} and DARWIN/XLZD\,\cite{Aalbers:2016jon}.

\acknowledgments

This work was supported by the National Science Foundation through grant \#2112803 as well as the Purdue Research Foundation.
We acknowledge the support of the Max Planck Society and like to thank the technicians of MPIK, Jonas Westermann and Michael Rei\ss felder, for their assistance.

\bibliography{manuscript}
\bibliographystyle{JHEP}

\end{document}